\pdfoutput=1
\documentclass[aps,twocolumn,superscriptaddress, prl]{revtex4-2}
\usepackage{graphicx}
\usepackage{dcolumn}
\usepackage{bm}
\usepackage{amsmath}
\usepackage{amssymb}
\usepackage{braket}
\usepackage[caption=false]{subfig}
\usepackage[colorlinks=true]{hyperref}
\usepackage{microtype}

\usepackage{physics}
\usepackage{mathrsfs}
\usepackage{comment}

\def\*#1{\mathbf{#1}}
\def\mc#1{\mathcal{#1}}
\def\mb#1{\mathbb{#1}}

\def\t#1{\text{#1}}

\def\mf#1{\mathfrak{#1}}

\def\bs#1{\boldsymbol{#1}}
\def\tt#1{\textit{#1.}|}

\begin{document}
\title{Topological transport of vorticity on curved magnetic membranes}

\author{Chau Dao}
\affiliation{Department of Physics and Astronomy and Bhaumik Institute for Theoretical Physics, University of California, Los Angeles, California 90095, USA}

\author{Ji Zou}
\affiliation{Department of Physics, University of Basel, Klingelbergstrasse 82, CH-4056 Basel, Switzerland}

\author{Eric Kleinherbers}
\affiliation{Department of Physics and Astronomy and Bhaumik Institute for Theoretical Physics, University of California,  Los Angeles, California 90095, USA}

\author{Yaroslav Tserkovnyak}
\affiliation{Department of Physics and Astronomy and Bhaumik Institute for Theoretical Physics, University of California, Los Angeles, California 90095, USA}

\begin{abstract}
     In this work, we study the transport of vorticity on curved dynamical two-dimensional magnetic membranes. We find that topological transport can be controlled by geometrically reducing symmetries, enabling processes absent from flat magnetic systems. To this end, the vorticity 3-current is constructed, which obeys a continuity equation immune to local disturbances of the magnetic texture and spatiotemporal fluctuations of the membrane. We show how electric current can manipulate vortex transport in geometrically nontrivial magnetic systems. As an illustrative example, we propose a minimal setup that realizes an experimentally feasible energy storage device and discuss its thermodynamic efficiency in terms of a \textit{vortexoelectric} counterpart of the thermoelectric figure of merit $ZT$.
\end{abstract}

\maketitle

\tt{Introduction}Much progress has been made both theoretically and experimentally in understanding, engineering, and driving topological magnetic textures~\cite{zang18,maekawa17,turnerRMP10,zhangJPCM20,maelandPRL23,cherepovPRL12,wangPRL22,dasguptaPRL20,ranaNATN23,schwartzPRR22,vogelPRB23,stepanovaNL22}. This has resulted in numerous proposals that exploit spin texture topology for technological applications, such as domain wall and skyrmion racetracks~\cite{parkinSCI08,blasingPIEEE20,tomaselloSR14,tomaselloIOP17,mullerNJP17}, energy storage~\cite{otxoaPRR21,tserkovnyakPRL18,jonesPRB20}, long-range signal transport~\cite{zouPRL20,zouPRB19,cornelissenNATP15,wangPRL19,raftreyPRL21,liuPRL20,tserkovnyakPRB20,tserkovnyakPRR19,zhangPRL20}, and quantum information processing~\cite{psaroudakiPRL21,xiaPRL23,zouARX22}. The utility of these spin structures is rooted in the metastability of topological excitations and the variety of ways to manipulate them~\cite{merminRMP79,stepanovPRL17,vogelPRB23, zhangJPCM20,chappertNATM10}. To foster the development of these technologies, it is crucial to innovate avenues to drive topological textures. Motivated by the interplay between geometry and topology~\cite{nakahara18, baez94, frankel11}, we seek a way to geometrically control topological transport. Previous works have investigated geometrical effects in curved low-dimensional magnetic systems, primarily focusing on energetic stabilization of topological spin textures~\cite{makarovSN22, makarovAM22, streubelIOP16, wangAP20,hernandezACSN20,shekaSMALL22}. These developments are spurred by advancements in fabrication and imaging techniques for complex magnetic structures~\cite{pachecoNATC17,hernandezACSN20,donnellyNATN22,donnellyNAT17}. 

In this Letter, we study the transport of vorticity on curved magnetic membranes. To this end, we demonstrate that magnetic textures on curved membranes exhibit \textit{topological hydrodynamics}, a transport theory for vortices~\cite{tserkovnyakJAP18}. This is governed by a robust continuity equation rooted in the homotopic properties of the magnetic order parameter space, rather than any structural symmetries. In contrast to previous studies focused on vortex transport in flat magnetic films~\cite{zouPRB19,zouPRL20,jonesPRB20,ahariPRB21}, this work leverages the geometry of curved magnetic films to reduce symmetries, thereby enabling processes that are otherwise ruled out on symmetry grounds. Notably, the chirality of the system plays a key role in allowing electric current to energetically bias vorticity injection on the membrane. 
A potential functionality of this physics is then illustrated by an experimentally feasible energy storage concept. 
Inspired by thermoelectricity, we derive the dimensionless \textit{vortexoelectric} figure of merit $\mc Z_v$ as the counterpart to the thermoelectric figure of merit $ZT$ and discuss the thermodynamic efficiency of the battery via this quantity.

\begin{figure}
    \centering
    \includegraphics[width=0.9\linewidth]{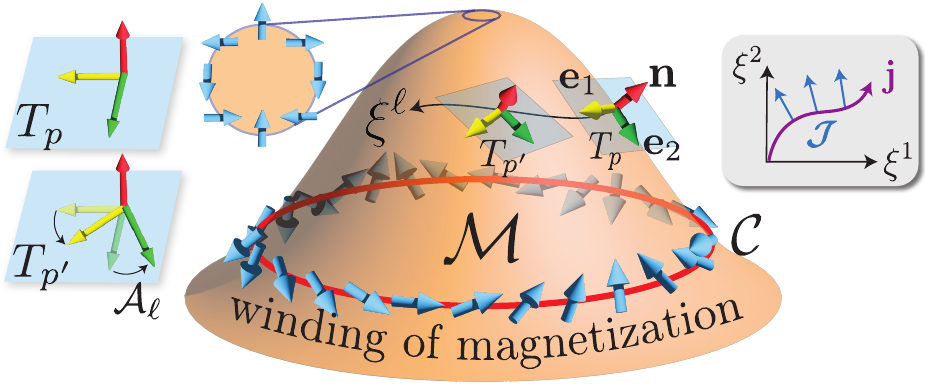}
    \caption{Depiction of manifold $\mc M$. The winding of the magnetization $\*m$ (blue arrows) on the contour $\mc C$ (red line) determines the vortex charge enclosed by $\mc C$. The local tangent planes $T_p$ and $T_{p'}$ are shown for points $p$ and $p'$, along with the local frame $\{\*e_1,\*e_2,\*n\}$. The leftmost tangent planes depict the gauge potential $\mc A_\ell$ at point $p'$, which captures changes of $\*e_1$ and $\*e_2$ along $\xi^\ell$. The inset shows vorticity flux $\bs{\mc J}$ being pumped transverse to a metallic wire (purple curve) carrying electric current density $\*j$.}
    \label{fig:manifold}
\end{figure}

\tt{Topological continuity equation}The magnetic membrane is a two-dimensional orientable manifold $\mc M$, parameterized by coordinates $\xi^1$ and $\xi^2$, with boundary $\partial\mc M$. $\mc M$ is embedded in the Euclidean space $\mb R^3$ from which it inherits the metric $g_{ij}$. At every point $(\xi^1,\xi^2)$ on $\mc M$ and for any time $t$, we identify a unit normal vector $\*n(t,\xi^1,\xi^2)$ and define unit vectors spanning the local tangent plane, $\*e_1(t,\xi^1,\xi^2)$ and $\*e_2(t,\xi^1,\xi^2)$. The orthonormal triad $\{\*e_1$, $\*e_2$, $\*n\}$ is the local frame. $\mc M$ may smoothly change over time, as long as its topology remains unchanged, i.e., $\mc M$ is not cut. Figure \ref{fig:manifold} depicts the simplest case in which $\mc M$ is homeomorphic to a closed disk.

The $\t U(1)$ gauge freedom in specifying the local frame, corresponding to simultaneous rotations of $\*e_1$ and $\*e_2$ about $\*n$, translates into the gauge potential~\cite{schwichtenbergARX19,kamienRMP02,zee03}
\begin{equation}\label{eq:Amu}
    \mc A_\mu = \*e_1\cdot\partial_\mu\*e_2,
\end{equation}
which is a smooth field describing changes of the local frame in space and time. Once the $\*e_1$ and $\*e_2$ vector fields are specified, the gauge is fixed. The spatial component of the gauge, $\mc A_i$, is the smooth connection on $\mc M$, which captures changes in $\*e_1$ and $\*e_2$ along $\xi^i$~\cite{kamienRMP02}. The gauge-independent field strength tensor stemming from $\mc A_\mu$ is
\begin{equation}\label{eq:fmunu}
    \mc F_{\mu\nu} = \partial_\mu\mc A_\nu- \partial_\nu \mc A_\mu.
\end{equation}
We can invoke the Mermin-Ho relation to express the field-strength tensor in terms of the surface normal as $\mc F_{\mu\nu} =\*n \cdot (\partial_\mu \*n \times \partial_\nu \*n)$, making apparent its gauge independence.
The ``electric" component, $\mc F_{0i}$, vanishes for static membranes. The ``magnetic" component, $\mc F_{ij}$, relates to the Gaussian curvature $\mc K$ by $\mc K=\mc F_{12}/\sqrt{g}$~\cite{kamienRMP02}. Here, $g$ is the determinant of the metric and the Levi-Civita tensor convention is $\epsilon^{012} = 1$. We make a convention in which Greek indices $\mu = 0,1,2\leftrightarrow t,\xi^1,\xi^2$ label spacetime coordinates and Latin indices $i = 1,2\leftrightarrow \xi^1,\xi^2$ label spatial coordinates, while repeated indices are summed over.

We assume the magnetic texture is described by a continuum coarse-grained vector field $\*m(t,\xi^1,\xi^2)$ realizing the map $\mc M \mapsto \mb R^3 $ at all times $t$. This description holds over a broad temperature range, from order to disorder. In the low-temperature (locally) ordered phase, $\*m$ is normalized by its $T=0$ value and the membrane can be either ferromagnetic or antiferromagnetic. $\*m$ is the local spin density in the former case, whereas in the latter case, $\*m$ is the local Néel order. In the high-temperature paramagnetic regime, $\*m$ may fluctuate in both magnitude and direction. Irrespective of any local fluctuations of $\*m$ or dynamics of $\mc M$, the field $\*m$ exhibits topological hydrodynamics governed by the continuity equation $\partial_\mu \mc J^\mu=0$, where
\begin{equation}\label{eq:vorticity_current}
        \mc J^\mu = \frac{\epsilon^{\mu\nu\rho}}{2\pi}\left[\*n\cdot\left(\nabla_\nu\*m\times\nabla_\rho\*m\right) -\frac{1}{2}\mc F_{\nu\rho}\*m_{\|}^2\right],
\end{equation}
written using the gauge-covariant derivative of $\*m$
\begin{equation}\label{eq:covariant_derivative}
        \nabla_\mu\*m \equiv (\partial_\mu m^a)\*e_a - \mc A_\mu (\*n\times\*m).
\end{equation}
Here, $\*m_\|$ is the projection of $\*m$ onto the local tangent plane. $\mc J^\mu = (\mc J^0, \bs{\mc J})$ is the gauge-independent vorticity 3-current, where $\mc J^0$ is the vorticity density and $\bs {\mc J}$ is the vorticity flux. In the absence of curvature and membrane dynamics, the frame can be made constant, so the vorticity 3-current reduces to $\mc J^\mu = \epsilon^{\mu\nu\rho}\*n\cdot(\partial_\nu\*m\times\partial_\rho\*m)/2\pi$~\cite{zouPRB19, jonesPRB20}. 

For flat systems, in the limit of an in-plane magnetic texture with fixed magnitude, the field homotopy defined on the boundary dictates the conserved integer vortex charge in the bulk~\cite{merminRMP79}. Whenever a vortex is moved across the boundary, there must be a commensurate change in the winding. Here, the topological continuity equation generalizes these ideas for curved systems with a magnetic texture that can traverse out of plane and fluctuate in magnitude. In contrast, the known skyrmion continuity equation~\cite{papanicolaouNPB91} does not derive from such a bulk-boundary correspondence. Hence, skyrmions can locally fluctuate in and out of existence if $|\*m|$ is allowed to vary~\cite{sitteJAP18}.
The continuity equation for vorticity is rooted in topology and is derived independent of the details of the system, thus being immune to any structural imperfections or anisotropies. However, the physical behavior of the system will be highly sensitive to these details~\cite{turnerRMP10,vitelliPRL04}. In this work, we will focus on systems in which magnetic vorticity may be the natural transport quantity. We consider magnetic membranes with easy-surface anisotropy, namely, there is a local hard anisotropy axis collinear with the surface normal~\cite{streubelIOP16,gaidideiPRL14} that endows the magnetic texture with an XY character~\cite{kosterlitzJPC73, turnerRMP10}.

\tt{Gauge-independent topological charge}To elucidate the bulk-boundary correspondence, we will construct the gauge-independent vortex density in terms of the magnetic winding. It is known that for flat magnetic films, the winding along a curve is $\*m_\|^2\partial_\ell\varphi/2\pi$~\cite{zouPRB19,jonesPRB20,tserkovnyakPRB20}, with $\ell$ the arclength. Following this structure, the winding on curved membranes generalizes to the gauge-covariant winding $\*m_\|^2D_\ell\varphi/2\pi \equiv \*m_\|^2(\partial_\ell\varphi - \mc A_\ell)/2\pi$. Here, $\varphi$ is the polar in-(tangent)-plane angle of $\*m_\|$ relative to $\*e_1$.  For $\varphi$ to be well-defined, we require $|\*m_\||>0$ everywhere on $\mc M$, except for isolated points. Invoking the generalized Stokes' theorem~\cite{stone09,nakahara18}, the gauge-independent topological charge on a patch $\mc S$ is
\begin{equation}\label{eq:Q}
    \mc Q = \frac{1}{2\pi}\int_{\partial\mc S}d\xi^i\,\*m_\|^2D_i\varphi = \int_\mc S  d\xi^1d\xi^2 \mc J^0.
\end{equation}
We see that the integrated winding around the boundary $\partial\mc S$ equals the vorticity density integrated over $\mc S$. The conserved density $\mc J^0$ is derived by the exterior derivative of the winding 1-form in the leftmost integral. We expand upon this construction in the Supplemental Material~\cite{supmat_dao23}. 

Even when specializing to the strong easy-surface limit where $|\*m_\||=1$, the topological charge $\mc Q$ is noninteger-valued if $\mc M$ has nonzero Gaussian curvature $\mc K$. In this limit, we can evaluate $\mc Q=\mc N - \int_{\mc S}dS\,\mc K/2\pi$ for a patch $\mc S$. This is the difference of an integer $\mc N$ that counts the number of vortices on $\mc S$ and a geometric background offset that spoils the discreteness of $\mc Q$. $\mc N$ is the $S^1$ winding number that connects $\mc Q$ to its homotopic roots.  A similar geometry-induced offset to the topological charge also appears when calculating the skyrmion number on curved surfaces~\cite{kravchukPRB16}.

\tt{Torsion-enabled pumping of vorticity}Having established a conserved topological charge, we now wish to control it. Suppose we wrap a static magnetic membrane with a metallic wire parameterized by unit tangent vector $\*v(\ell)$, with $\ell$ the arclength, and induce electric current flow in the wire. Following the approach developed in Refs.~\cite{zouPRB19, jonesPRB20}, we construct a torque acting on a smooth magnetic texture, which energetically biases vorticity injection transverse to the wire, as depicted in the inset of Fig.~\ref{fig:manifold}. This engenders the \textit{vortexoelectric effect}. To enable this process, we use the local torsion of the wire, $\mf T(\ell)= \*v\cdot(\*a \times\partial_\ell\*a)$, to geometrically reduce symmetries~\cite{stone09}. The torsion is the helical winding of the principal normal vector $\*a(\ell)=\partial_\ell\*v/|\partial_\ell\*v|$ along the wire, and is zero for curves on flat surfaces. This means that for flat magnetic systems, this process would be ruled out on symmetry grounds. Notably, torsion is a pseudoscalar, so $\mf T\*n$ is a pseudovector with identical spatial properties to magnetization which can be substituted for $\*M$ in Refs.~\cite{zouPRB19, jonesPRB20}.

The gauge-independent torque is constructed phenomenologically on general symmetry grounds so that the work done on the magnetic texture is proportional to the vorticity flow across the wire. Focusing, for simplicity, on the low-temperature limit where $|\*m|=1$, we require a torque orthogonal to $\*m$. A torque (per unit length) that satisfies this is~\cite{supmat_dao23}
\begin{equation}\label{eq:tnt}
    \bs\tau = \frac{\zeta j}{\pi}(\mf T\*n\cdot\*m)\left[\nabla_\ell \*m+\*n\,\partial_\ell(\*n\cdot\*m)\right].
\end{equation}
Here, $j = \*j\cdot\*v$ is the electric current density, and $\zeta$ is a phenomenological parameter characterizing the geometry-enabled dissipative coupling of electric and vortex dynamics. Importantly, this torque is only permitted in the presence of spin-orbit coupling. This reminds us of the chirality-induced spin selectivity effect, which arises from the coupling of the electron linear momentum to spin degrees of freedom in chiral materials~\cite{naamanJPCL12,dalumNL19}. Furthermore, in the limit of strong spin-orbit coupling, dimensional analysis suggests $\zeta\sim\hbar w\lambda_F^2/e$, where $\lambda_F$ is the Fermi wavelength, $e$ is the positive elementary electric charge, and $w$ is the width of the wire.

The work done on the magnetic texture is
\begin{align}\label{eq:deltaW}
    \delta W=\int d\ell\,dt \,\vb*\tau \cdot (\* m \times \partial_t \* m) = \zeta \mf Tj\,\delta\mathcal{Q},
\end{align}
where $\delta \mc Q = \int dtd\ell\,\*v\cdot(\bs{\mc J}\times\*n)$ is the vorticity flow across the wire. The effective vortex chemical potential is $\mu\equiv\delta W/\delta Q= \zeta \mf T j$. In the high-temperature paramagnetic regime, a linear relation $\mu\propto j$ should still hold, although thermal fluctuations of $|\*m|$ will renormalize the prefactor.

\begin{figure}
    \centering
    \includegraphics[width=\linewidth]{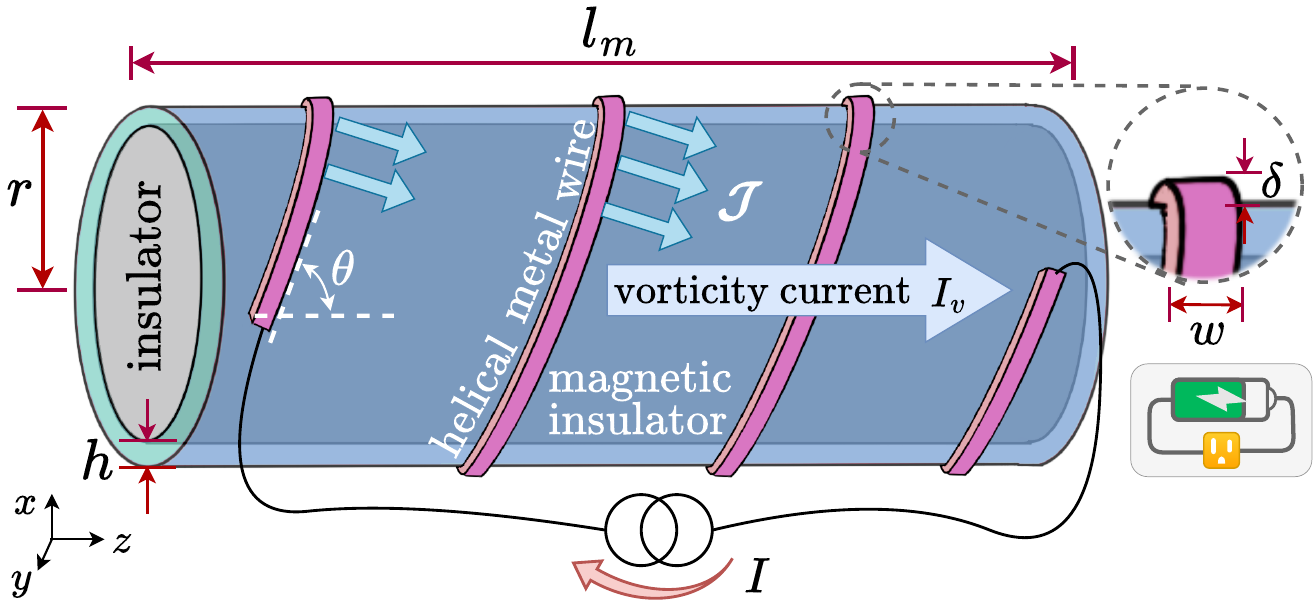}
    \caption{Schematic of a minimal setup for geometrically controlled vortex transport. A metallic wire is wrapped around a cylindrical magnetic insulator membrane as a helix. An applied electric current $I$ induces vorticity flux $\bs{\mc J}$ transverse to the wire, resulting in vorticity current $I_v$ along $z$. The side panel indicates that this system realizes a battery.}
    \label{fig2}
\end{figure}

\tt{Vortex circuit elements}Fig.~\ref{fig2} illustrates a possible setup where torsion gives rise to pumping of vorticity. Here, a magnetic insulating membrane (which can either be ferro- or antiferromagnetic) of thickness $h$ and length $l_m$ wraps around a cylindrical insulating core. A metal wire of width $w$ and thickness $\delta$ wraps around the cylindrical magnetic membrane of radius $r$ as a uniform helix with helix angle $\theta$. Systems with similar geometry, in the form of rolled magnetic membranes, have been fabricated~\cite{streubelNATC15}. The uniform helix has a constant torsion $\mf T = \sin(2\theta)/2r$, allowing electric current flow in the wire to drive a vorticity flux $\bs{\mc J}$, which we assume is transverse to the wire. For this setup, the vortex current will flow along the cylinder in the same direction as the electrical current flow, irrespective of the sign of the torsion $\mf T$~\footnote{Electrical current flow in the $z$ direction will generically drag an additional vortex current, since vortices and antivortices are not related by structural symmetries, thus having distinct physical properties. For example, antivortices could be immobile while vortices are not. We expect this drag effect to be insensitive to the sign of the torsion $\mf{T}$, as well.}.

$\bs {\mc J}$ can be decomposed into components orthogonal and parallel to the $z$ axis. The former accumulates winding along $z$, which may unwind at the ends of the cylinder. The latter, on the other hand, builds up winding azimuthally, which is energetically protected by the easy-surface anisotropy. We are interested in the vortex current flowing in the $z$ direction, $I_v = 2\pi r|\boldsymbol{\mc J}|\sin\theta$, which is driven by the vortex motive force $I\mf R$ generated by the electric current. Here, 
\begin{equation}
    \mf R = \frac{\zeta \mf Tl_m}{2\pi rw\delta}\tan\theta
\end{equation}
is the vortexoelectric drag coefficient, which can be understood as the analog to the Seebeck coefficient~\cite{supmat_dao23}. Different from our previous works on energy storage using topological spin textures~\cite{tserkovnyakPRL18,jonesPRB20}, $\mf R$ is unique to the nontrivial geometry of this setup and is absent from flat counterparts. Upon substitution of $\mf T = \sin(2\theta)/2r$, we find $\mf R\propto \sin^2\theta$. In the linear response, $I_v\propto \mf R$, so $\theta=\pi/2$ maximizes the vorticity flow. 

The membrane behaves like a series $R_vC_v$ circuit in response to nonzero vortex flow, exhibiting an effective vortex resistance $R_v$ and effective winding capacitance $C_v$.  As dictated by the bulk-boundary correspondence, the vortex current ``winds up" the magnetic texture, thereby storing exchange energy. The stiffness of the magnetic texture engenders $C_v$. On the other hand, $R_v$ can arise due to Gilbert damping, defects, and vortex-antivortex collisions. Following Ref.~\cite{jonesPRB20}, we estimate $C_v$ and $R_v$ by exploiting the duality between the XY magnet and two-dimensional electrostatics~\cite{kosterlitzJPC73}. We find
\begin{equation}
    C_v=\frac{1}{A}\frac{r}{2\pi hl_m},~~~R_v=\frac{1}{\sigma_v}\frac{l_m}{2\pi r},
\end{equation}
where $\sigma_v^{-1}$ is the vortex resistivity and $A$ is the magnetic stiffness~\cite{supmat_dao23}. With the circuit elements $\mf R, R_v$, and $C_v$ in hand, we set out to construct topological circuits~\cite{jonesPRB20}.

\begin{figure}
    \centering
    \includegraphics[width=\linewidth]{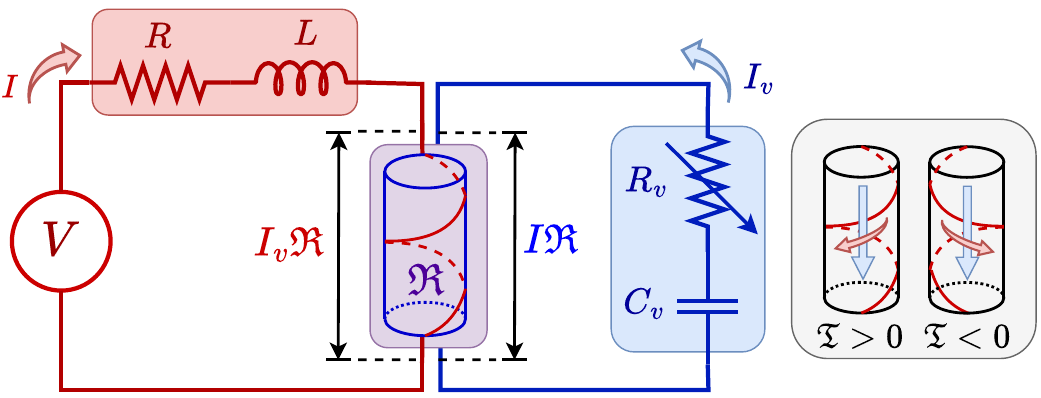}
    \caption{Schematic of the vorticity (blue) and electric (red) circuits, which are coupled through $\mf R$ (purple). $\mf R$ gives rise to an effective vortex motive force $I\mf R$ on the vorticity circuit and, reciprocally, an electromotive force $I_v\mf R$ on the electrical response. $R_v$ is tunable, allowing switching between vortex conducting and insulating regimes. The side panel depicts setups for positive and negative $\mf T$, with both electrical and vortex currents flowing downwards.}
    \label{fig:circuits}
\end{figure}

\tt{Coupled topological circuits}The setup we have been discussing can be described by coupled vorticity and electric circuits, which are depicted in Fig.~\ref{fig:circuits}. The applied electric current $I$ in the wire supplies an effective vortex motive force $I\mf R$ to the vorticity circuit. This results in build-up of winding and an effective vortex voltage $V_v = -\mc Q/C_v$. The backaction of vortex dynamics on the electrical response induces an electromotive force $I_v\mf R$ on the electric circuit, which is written down by invoking Onsager reciprocity. 
Note that, like ordinary charge, vorticity is even under time reversal. Kirchhoff's law for the coupled electrical and vorticity circuits is thus
\begin{equation}\label{eq:coupledcircuit}
    \mqty(V \\ V_v)=\mqty(R+L\frac{d}{dt} & -\mf R \\ -\mf R & R_v) \mqty(I \\ I_v).
\end{equation}
Here, $V$ is the electric voltage, $L$ is the self-inductance, and $R$ is the electrical resistance. The resistance matrix is symmetric as dictated by Onsager reciprocity, and positive-definite according to the second law of thermodynamics~\cite{onsagerPR31,gyarmati70}. The latter constraint enforces $0<\xi<1$, where $\xi \equiv \mf R^2/RR_v$ parameterizes the relative strength of the off-diagonal to the diagonal elements of the resistance matrix.

Fourier transforming Eq.~(\ref{eq:coupledcircuit}) into the frequency domain, we find the effective impedance is
\begin{equation}\label{eq:Z}
Z(\omega)\equiv \frac{V(\omega)}{I(\omega)}=R+i\omega L -  \frac{i\omega C_v \mf R^2 }{1+i\omega C_vR_v}.
\end{equation}
Similar to conventional RC circuits, here, $\tau= R_vC_v = (4\pi^2Ah\sigma_v)^{-1}$ is the time scale for loading and discharging vortices from the magnetic texture. In the high-frequency regime, $\omega \gg 1/\tau$, the last term in Eq.~(\ref{eq:Z}) approximates $-\mf R^2/R_v.$ The vorticity circuit reduces the effective resistance of the electrical circuit. In the low-frequency limit, $\omega \ll 1/\tau$, the vorticity circuit acts like an inductor with effective negative inductance $L_v = -C_v\mf R^2$. Impedance measurements of the circuit in the low-frequency regime could pave a way to probe the strength of the coupling between vortex and charge currents. Similar impedance measurements on helical-spin magnets have been performed to characterize the current-driven dynamics of spin-helix structures~\cite{yokouchiNAT20}.

\tt{Energy storage and efficiency}In addition to providing a means to measure $\zeta$, the setup depicted in Fig.~\ref{fig2} may also function as a battery. Operation of the battery requires a mechanism to switch the vortex conductivity between the conducting and insulating regimes, allowing the battery to alternate between (dis)charging and storing energy, respectively. The vortex transport parameters could be very sensitive and may be modulated, for example, by heating and cooling the magnet~\cite{kosterlitzJPC73,jonesPRB20}.

To charge the battery, we electrically bias vortex flow along $z$, building up azimuthal spin winding so the magnet accumulates exchange energy. Discharging the battery is the reverse process wherein a vortex current induces an electromotive force on the electric circuit, which we may extract as energy. We store the exchange energy by lowering the vortex conductivity, so vortex transport parameters enter the insulating regime. Once in the insulating regime, the Landau criterion governs the amount of winding we can stabilize since the magnetic bulk cannot host an arbitrarily sharp texture~\cite{soninAIP10,tserkovnyakPRL18}. The easy-surface anisotropy $({\sim} K)$ protects the topological spin texture by energetically preventing ``phase-slip" events during which the magnetic order parameter unwinds~\cite{kimPRB16}. Thus, easy-surface anisotropy determines the maximal energy storage capacity, which saturates when winding texture energy $[{\sim} A(D_\ell \varphi)^2]$ is comparable to $K$.

The charging and discharging efficiencies may be used to characterize the battery. In the vortex conducting regime, we charge the battery relative to its ground state by supplying a dc electric current $I_0$ for duration $\tau$. By tuning $R_v$, we can switch to the vortex insulating regime to store the energy in the winding capacitor. The charging efficiency $\eta_c$ is the ratio of the stored energy to the total energy supplied by the electric circuit. We extract the stored energy by connecting the battery to a load resistor $R_L$, then switching back to the vortex conducting regime to discharge. The discharging efficiency $\eta_d$ is the ratio of energy consumed by $R_L$ to the energy leaving the winding capacitor. 

Neglecting the self-inductance $L$, the efficiencies are 
\begin{equation}
    \eta_c = \frac{1}{2}\frac{(1-e^{-1})^2}{{\mc Z_v}^{-1}+e^{-1}},~~~\eta_d=\frac{1-\gamma}{1+(\mathcal Z_v\gamma)^{-1}},
\end{equation}
written with $\gamma = R_L/(R_L+R)$ and $\mc Z_v\equiv \xi/(1-\xi)$, where $\xi = \mf R^2/RR_v$. Here, we define the \textit{vortexoelectric} figure of merit $\mc Z_v$ by analogy to the thermoelectric figure of merit $ZT$~\cite{majumdarSCI04,kovalevSSC10,bauerPRB10}. Whereas for the thermoelectric effect, heat and charge currents are coupled, in this setting, we cross-couple vortex and electric currents. Since $\mc Z_v$ is a monotonic function of $\xi$, optimizing the system geometry to maximize $\xi\sim adw\lambda_F^2\sin^4\theta\cos\theta/hr^3\delta$ maximizes $\mc Z_v$ and, hence, the efficiencies. Here, $d$ is the electron mean free path and $a$ is the lattice spacing. $\mc Z_v$ is improved by decreasing $r$, thinning the membrane and the metal wire by decreasing $\delta$ and $h$, or enlarging the metal-magnet interface by increasing $w$. The optimal helix angle is $\theta \approx 63^\circ$, which balances maximizing $I_v$ and minimizing energy lost due to Joule heating. In the maximal efficiency limit, $\mc Z_v\rightarrow\infty$, the efficiencies simplify to $\eta_c = (e-1)^2/2e$ and $\eta_d = 1-\gamma$ for (dis)charging times of $t=\tau$. Furthermore, in the short charging time limit of $t/\tau\rightarrow0$, while still having $\mc Z_v\rightarrow \infty$, the charging efficiency saturates as $\eta_c\rightarrow 1$.

\tt{Discussion}In this work, we developed topological hydrodynamics for vortices on curved membranes and proposed an avenue to use geometric properties of the system, in particular chirality, as a tunable handle to bias vortex transport. To achieve this, we phenomenologically introduced a torque that uses geometric torsion to enable electric current-induced vortex pumping. With these building blocks, we analyzed a minimal setup which is described by an effective coupled electrical-vorticity circuit and realizes a feasible energy storage concept. We derive the vortexoelectric figure of merit $\mc Z_v$ and use this quantity to characterize the thermodynamic efficiency of the battery.

For future works, we may extend our formalism to manifolds with different topologies, such as the Möbius strip. These setups may potentially be achieved by modern fabrication techniques, which have been employed to manufacture complex magnetic structures~\cite{shekaSMALL22,streubelNATC15,phatakNL14,skoricNL20}. It is important to note that the topological continuity equation we formulated applies not only to magnetic systems but also more broadly to any system with a vectorial order parameter. In particular, these concepts can be directly applied to describe the transport of nematic disclinations in liquid crystals, 
a topic of much recent interest~\cite{vafaPRE24, vafaJPCM23,jiangLC23,duclosSCI20,evertsPRX21,sandfordNATC20}. Another possible direction would be to expand beyond vortex circuit elements and develop logic elements based on this physics. 

\begin{acknowledgments}
\textit{Acknowledgments.} We thank Denys Sheka and Se Kwon Kim for insightful discussions. This work was primarily supported by the U.S. Department of Energy, Office of Basic Energy Sciences under Grant No. DE-SC0012190. J.Z. acknowledges the support of the Georg H. Endress Foundation.
\end{acknowledgments}

\bibliography{ms.bib}
\end{document}


\preprint{APS/123-QED}

\title{Supplemental Material for \\ ``Topological transport of vorticity on curved magnetic membranes"}

\author{Chau Dao}
\affiliation{Department of Physics and Astronomy and Bhaumik Institute for Theoretical Physics, University of California, Los Angeles, California 90095, USA}

\author{Ji Zou}
\affiliation{Department of Physics, University of Basel, Klingelbergstrasse 82, CH-4056 Basel, Switzerland}

\author{Eric Kleinherbers}
\affiliation{Department of Physics and Astronomy and Bhaumik Institute for Theoretical Physics, University of California, Los Angeles, California 90095, USA}

\author{Yaroslav Tserkovnyak}
\affiliation{Department of Physics and Astronomy and Bhaumik Institute for Theoretical Physics, University of California, Los Angeles, California 90095, USA}

\maketitle

In this Supplemental Material, we provide (i) the construction of the gauge-independent winding density, (ii) the differential geometric formulation of topological hydrodynamics, (iii) a derivation of the torsion of a uniform helix,  (iv) a discussion on how the phenomenological torque is constructed, and (v) a discussion on how $R_v, C_v$, and $\mf R$ are estimated.

\subsection{(i) Gauge-independent winding density}

In this section, we construct the gauge-independent winding density, discussed in the section ``Gauge-independent topological charge" of the main text. This winding density is the starting point for formulating topological hydrodynamics of vortices on curved magnetic membranes. Let us consider a curved dynamical two-dimensional magnetic membrane. The membrane is a two-dimensional orientable manifold $\mc M$ with boundary $\partial\mc M$ and is parameterized by global coordinates $\xi^1$ and $\xi^2$. $\mc M$ is embedded in the Euclidean space $\mb R^3$ from which it inherits the metric $g_{ij}$. At every point $(\xi^1,\xi^2)$ on $\mc M$ and for any time $t$, we identify a unit normal vector $\*n(t,\xi^1,\xi^2)$ and define unit vectors spanning the local tangent plane, $\*e_1(t,\xi^1,\xi^2)$ and $\*e_2(t,\xi^1,\xi^2)$. $\{\*e_1$, $\*e_2$, $\*n\}$ form an orthonormal triad and is the local frame. We make the convention in which Greek indices $\mu = 0,1,2\leftrightarrow t,\xi^1,\xi^2$ label spacetime coordinates, Latin indices $i = 1,2\leftrightarrow \xi^1,\xi^2$ label spatial coordinates, while repeated indices are summed over. 

The magnetic texture is described by the continuum coarse-grained vector field $\*m(\*r,t)$. In the low-temperature ordered phase, $\*m$ exhibits only orientational dynamics and is normalized by its $T=0$ value. On the other hand, in the high-temperature paramagnetic regime $\*m$ can fluctuate in both magnitude and direction. To construct the gauge-independent winding density, it is useful to first specialize to the ordered phase and take the strong easy-surface limit, in which $|\*m|=1$ and $\*m$ lies fully within the local tangent plane. $\*m$ can be written in the local frame as
\begin{equation}
    \*m(t,\xi^1,\xi^2) = \*e_1\cos\varphi + \*e_2 \sin\varphi,
\end{equation}
where $\varphi$ is the in-(tangent)-plane angle of $\*m$ relative to $\*e_1$. For a closed loop $\partial\mc S$ on $\mc M$, $\varphi$ realizes the map $S^1\mapsto S^1$. For flat manifolds, the winding number would simply be the degree of this map and would be given by $\mc Q = \int_{\partial S} d\xi^i\,\*m_\|^2 \partial_i\varphi/2\pi$ \cite{zouPRB19,jonesPRB20}. However, different from winding in flat geometries, the local frame cannot be made constant, and will generally vary over $\mc M$. This means that $\partial_i\varphi$ would not be gauge independent. To define the winding, we must be able to compare the phase $\varphi$ of the magnetization at different points on the curved manifold in a gauge-independent fashion. Taking the derivative of $\*m$ and projecting onto the local basis vectors, we find that
\begin{subequations}\label{eq:projections}
    \begin{align}
        \*e_1\cdot\partial_i\*m &= -\sin\varphi\,[\partial_i\varphi - \*e_1\cdot \partial_i\*e_2], \\
        \*e_2\cdot\partial_i\*m &=  \cos\varphi\,[\partial_i\varphi - \*e_1\cdot\partial_i\*e_2].
    \end{align}
\end{subequations}
Both expressions yield the same bracketed expression. The first term in the brackets is the winding of the in-plane angle and the second term is the the winding of the phase that is induced by a changing frame. Formally, the second term in the brackets is the connection on the manifold,
\begin{equation}\label{eq:gauge_pot}
    \mc{A}_i \equiv \*e_1\cdot\partial_i\*e_2,
\end{equation}
which captures the changes in $\*e_1$ and $\*e_2$ along $\xi^i$. If the manifold is dynamic, the local frame will be time-dependent and we can define $\mc A_0 = \*e_1\cdot\partial_t\*e_2$, which tracks changes of the local frame in time. Together with the connection, we can define $\mc A_\mu = \*e_1\cdot\partial_\mu\*e_2$ as the gauge potential which stems from the U(1) gauge freedom corresponding to rotations of $\*e_1$ and $\*e_2$ about $\*n$. The term in the brackets of Eq.~\ref{eq:projections} is the gauge-covariant derivative of $\varphi$, which we define as 
\begin{equation}
    D_i\varphi \equiv \partial_i\varphi - \mc A_i.
\end{equation}
This construction of $D_i\varphi$ closely follows that done in Ref.~\cite{kamienRMP02}. Having constructed $D_i\varphi$, let us relax the constraint taken earlier, in which we specialized to the low-temperature ordered phase and took the strong-easy surface limit. We now allow $\*m$ to traverse out of the local tangent plane and fluctuate in both magnitude and direction. For flat systems, it is known that the winding density along a curve parameterized is $\*m_\|^2\partial_\ell\varphi/2\pi$ \cite{zouPRB19}. We can generalize this result for curved systems by making the substitution $\partial_i\varphi\rightarrow D_i\varphi$ so that the covariant winding is
\begin{equation}
    d\xi^i\,\frac{1}{2\pi}\*m_\|^2D_i\varphi = d\xi^i\,\frac{1}{2\pi}\*m_\|^2(\partial_i\varphi - \mc A_i).
\end{equation}
The topological charge $\mc Q$ enclosed on a patch $\mc S$ of the manifold is consequently 
\begin{equation}
    \mc Q = \int_{\partial \mc S} \rho_w = \frac{1}{2\pi}\int_{\partial S} d\xi^i\*m_\|^2D_i\varphi,
\end{equation}
where $\rho_w$ is the gauge-independent winding 1-form. The relation between $\mc Q$ and the winding density is given by Eq.~(5) in the main text. Thus, we have constructed the winding density for a curved magnetic membrane and connected it to the gauge-independent topological charge $\mc Q$.

\begin{figure} [b]
    \centering
    \includegraphics[width = 0.75\linewidth]{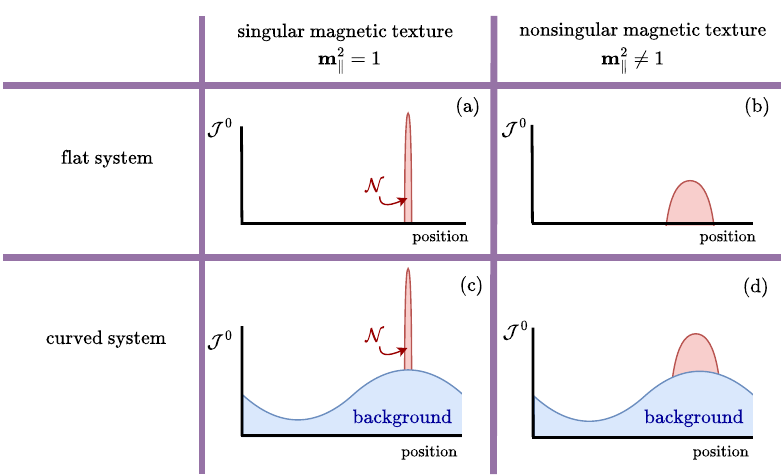}
    \caption{Cartoon depiction of the vorticity density for flat and curved magnetic membranes, and for singular and nonsingular magnetic textures. Panel (a) shows that for a flat magnetic system, taking the strong easy-plane limit in which $\*m_\|^2=1$ results in the magnetic texture becoming singular, and the vorticity density is a $\delta$-function at the location of the vortex. Panel (b) shows that relaxing the strong easy-plane limit and allowing the magnetic texture to be nonsingular results in broadening of the delta function. Panels (c) and (d) show that nonzero curvature adds a background offset to the vorticity density.}
    \label{fig:suppfig1}
\end{figure}

It is advantageous to be able to write the winding density $\rho_w$ in terms of $\*m$ rather than the in-plane angle $\varphi$. Doing so allows us to describe the topological charge density using a nonsingular vector field $\*m$, rather than the $\phi$ field which has singularities when vortices are present. To this end, we define the covariant derivative of $\*m$ as 
\begin{equation}
    \nabla_\mu\*m \equiv (\partial_\mu m^a)\*e_a - \mc A_\mu\*n\times\*m. \label{eq:covariant_derivative}
\end{equation}
This is identical to Eq.~(4) of the main text. Here, $\mc A_\mu = \*e_1\cdot\partial_\mu\*e_2$ can be understood as a gauge potential that tracks changes in the local frame in space as well as time. The covariant derivative of a vector in $\mb R^3$ at any given point on $\mc M$ will result in a vector lying in the local tangent plane. Using this covariant derivative, we can express the winding 1-form in terms of $\*m$ as
\begin{equation}
    \rho_w = d\xi^i \frac{1}{2\pi}\*n\cdot(\*m\times \nabla_i\*m),
\end{equation}
which is the integrand of Eq.~(13) in the main text.
Finally, we elaborate on a point made in the last paragraph of the section ``Gauge-independent topological charge" regarding the quantization of $\mc Q$. Generally, when there is nonzero Gaussian curvature, even when taking the strong-easy surface limit, $\mc Q$ is noninteger valued. A cartoon depiction of the vorticity density $\mc J^0$ for flat and curved membranes, with and without taking the easy-surface limit, is depicted in Fig.~\ref{fig:suppfig1}. Note that $\mc J^0$ is given by Eq.~(3) of the main text.

Taking the strong easy-surface limit, the topological charge on a patch $\mc S$ (upon picking a smooth gauge) of the membrane can be evaluated as 
\begin{equation}
    \mc Q = \mc N - \frac{1}{2\pi}\int_\mc S d\xi^1d\xi^2\sqrt{g} \mc K,
\end{equation}
where $\mc K$ is the local Gaussian curvature and $g$ is the determinant of the metric. We see that the topological charge $\mc Q$ is the difference of an integer $\mc N$, which counts the number of vortices on $\mc S$, and a geometrical background offset that spoils the discreteness of $\mc Q$. The effect of this background offset is depicted in panels (c) and (d) of Fig.~\ref{fig:suppfig1}. When the strong-easy surface limit is taken, we find that the vorticity density sharply peaks at the location of the vortex. In the $|\*m_\||=1$ limit, for flat membranes, the vorticity is a $\delta$-functions, and for curved membranes, the vorticity is the sum of the background offset and a $\delta$-function. However, by relaxing the strong easy-surface limit, we allow for a nonsingular magnetic texture and the vorticity density broadens, as shown in panels (b) and (d).

\subsection{(ii) Formulation of topological hydrodynamics}

In this section, we start from the covariant winding density, which was derived in the previous section, and construct the vorticity 3-current $\mc J^\mu$. We first specialize to the static manifold, in which the local frame has no time dependence, to construct the expression for $\mc J^\mu$. We will then show that $\mc J^\mu$ satisfies the topological continuity equation $\partial_\mu \mc J^\mu=0$, even in the presence of local fluctuations of $\*m$ or spatiotemporal fluctuations of the magnetic membrane. To do so, we will formulate topological hydrodynamics using differential forms. The topological charge on a patch $\mc S$ of the manifold $\mc M$ is 
\begin{equation}\label{eq:tangent_winding}
    \mc Q = \int_{\partial\mc S} \rho_w =\frac{1}{2\pi} \int_{\partial \mc S}d\xi^i \*n\cdot(\*m\times\nabla_i\*m),
\end{equation}
which is defined as the integral of the winding 1-form $\rho_w$ over the boundary of the patch, $\partial \mc S$. After defining the winding 1-form, we invoke the generalized Stoke's theorem, $ \int_{\partial\mc S} \rho_w= \int_{\mc S} d\rho_w$, to define the vorticity 2-form $\rho_v$ as the exterior derivative of $\rho_w$. We have that
\begin{align}
    \begin{split}
    \rho_v \equiv d\rho_w &= d\xi^i\wedge d\xi^j\frac{1}{2\pi}\partial_i\left[\*n\cdot\left(\*m\times\nabla_j\*m\right)\right]\\
    &= d\xi^i\wedge d\xi^j\frac{1}{2\pi}\*n\cdot\left(\nabla_i\*m\times\nabla_j\*m-\frac{1}{2}\mc F_{ij}\*m_\|^2\right)\\
    &= d\xi^1\wedge d\xi^2 \left[\frac{\epsilon^{0ij}}{2\pi}\*n \cdot\left(\nabla_i\*m \times\nabla_j\*m - \frac{1}{2}\mc F_{ij}\*m_\|^2\right)\right].
    \end{split}
\end{align}
Here, $d\equiv d\xi^i\wedge \partial_i$ is the exterior derivative and $\mc F_{ij}=\partial_i\mc A_j - \partial_j\mc A_i$ is the gauge-independent curvature field which arises from the connection $\mc A_i$ on $\mc M$. The curvature field $\mc F_{ij}$ is related to the Gaussian curvature $\mc K$ by $\mc K = \epsilon^{0ij}\mc F_{ij}/2\sqrt{g}$. Here, we use the Levi-Civita convention that $\epsilon^{012}=1.$ Furthermore, $\mc F_{ij}$ is the ''magnetic" component of the field strength tensor $\mc F_{\mu\nu}$ given by Eq.~(2) of the main text. We can also identify the function in the brackets as the temporal component of the vorticity 3-current $\mc J^\mu$, the vorticity density
\begin{equation}\label{eqsup:vorticitydensity}
\mc J^0 = \frac{\epsilon^{0ij}}{2\pi}\*n \cdot\left(\nabla_i\*m \times\nabla_j\*m - \frac{1}{2}\mc F_{ij}\*m_\|^2\right).
\end{equation}

For any change of the charge $\mc Q$ on $\mc S$, we expect to find a vorticity flux $j_v$ through the boundary $\partial S$, which changes the winding along the boundary. That is, we wish to find $j_v$ satisfying the continuity equation
\begin{equation}\label{eqsup:continuity}
\partial_t\rho_v + d \star j_v = 0.
\end{equation}
In this expression, we have used the Hodge star map ``$\star$", which maps a $p$-form in $d$-dimensional space to its Hodge dual, a $(d-p)$-form. The mapping is defined to be \cite{stone09}
\begin{equation}\label{eqsup:hodge}
    \star d\xi^{i_1}\ldots d\xi^{i_p} = \frac{1}{(d-p)!}\sqrt{g}g^{i_1j_1}\ldots g^{i_pj_p}\epsilon_{j_1\ldots j_p j_{p+1}\ldots j_d}d\xi^{j_{p+1}}\ldots d\xi^{j_d}.
\end{equation} 
The first term in Eq. (\ref{eqsup:continuity}), $\partial_t\rho_v$, is the time derivative of the vorticity 2-form, and the second term $d \star j_v$ is the divergence of the vorticity flux 1-form. To construct $j_v$, we start with the relationship between vorticity and winding, which states that $\rho_v = d\rho_w$. Taking the time derivative, we get 
\begin{equation}
    \partial_t \rho_v - \partial_t d\rho_w = 0.
\end{equation}
For any $p$-form $\omega$, applying the Hodge star map twice yields $\star\star\omega = (-1)^{d-p}\omega$. Using this property, and commuting $d$ and $\partial_t$, we rewrite this equality as
\begin{equation}
    \partial_t\rho_v +  d \star (\star\partial_t\rho_w) = 0.
\end{equation}
Now, we can identify the vorticity flux 1-form as 
\begin{align}
j_v=\star\partial_t\rho_w +\star d f,
\end{align}
where we have the freedom to add the term $\star d f$ with $0$-form $f$, because it has zero divergence, $d\star\left(\star d f\right)=0$. To agree with the flat-geometry case~\cite{zouPRB19,jonesPRB20}, we choose 
\begin{align}
    f=-\frac{1}{2\pi}\*n\cdot(\*m\times \partial_t\*m).
\end{align}
To calculate $df=d\xi^i\partial_i f$, we use the identity  \begin{align}
\partial_i \left[\vb{n}\cdot \left(\vb{a}\times\vb{b}\right)\right]=\vb{n} \cdot \left(\nabla_i \vb{a}\times\vb{b}\right)+\vb{n} \cdot\left( \vb{a}\times\nabla_i \vb{b}\right),\label{eqsup:identity}
\end{align}
where $\*n$ is the surface normal vector and $\*a$ and $\*b$ can be any vector in $\mb R^3$. This can be derived from a straightforward application of Eq.~\eqref{eq:covariant_derivative}.
Then, it follows that
\begin{align}
\begin{split}
    j_v &= \frac{1}{\pi}\star\left[d\xi^i\*n\cdot(\partial_t\*m\times\nabla_i\*m)\right]\\
        &= d\xi^k \frac{1}{\pi}\*n\cdot(\partial_t\*m\times\nabla_i\*m)\sqrt{g}g^{ij}\epsilon_{jk},
\end{split}
\end{align}
with the contravariant components
\begin{equation}
    j_v^i = \frac{1}{\sqrt g\pi}\*n\cdot(\partial_t\*m\times\nabla_j\*m)\epsilon^{ji}.
\end{equation}
Taking the divergence of $j_v$, we find that
\begin{align}\label{eqsup:vorticityflux}
    \begin{split}
        d \star j_v &= d\xi^1\wedge d\xi^2 \partial_i (\sqrt{g} j_v^i)\\
        &=d\xi^1\wedge d\xi^2\partial_i\left[\frac{1}{\pi}\*n\cdot(\partial_t\*m\times\nabla_j\*m)\epsilon^{ji}\right]\\
        &= d\xi^1\wedge d\xi^2 \partial_i \mc J^i.
    \end{split}
\end{align}
In the final line, we identify the term in the brackets with the vectorial vorticity flux $\bs{\mc J}$. The scalar vorticity density and the vectorial vorticity flux can be combined into a vorticity 3-current $\mc J^\mu = (\mc J^0, \boldsymbol{\mathcal{J}})$, where $\mc J^0$ is the vorticity density and $\boldsymbol{\mathcal J}$ is the vorticity flux. Using Eqs. (\ref{eqsup:vorticitydensity}) and (\ref{eqsup:vorticityflux}), we see that $\mc J^\mu$ is given by 
\begin{align}\label{eqsup:vorticity_current}
     \mc J^\mu =\frac{\epsilon^{\mu\nu\rho}}{2\pi}\left[\*n\cdot\left(\nabla_\nu\*m\times\nabla_\rho\*m \right)-\frac{1}{2}\mc F_{\nu\rho}\*m_\|^2\right].
\end{align}
$\mc J^0$ is related to $\rho_v$ by $\rho_v = \mc J^0d\xi^1d\xi^2$ and $\mc J^i$ is related to $j_v$ by $j_v = \mc J^ig_{ij}d\xi^j/\sqrt{g}$. This expression for the vorticity 3-current is given by Eq.~(3) in the main text. Here, $\mc F_{\nu\rho} = \partial_\nu \mc A_\rho - \partial_\rho \mc A_\nu$ can be understood as the gauge-independent field strength tensor, analogous to $\mc A_\mu = \*e_1 \cdot\partial_\mu \*e_2$ being understood as the gauge potential. The ``magnetic" component of the field strength tensor is the curvature field $\mc F_{ij}$. The topological continuity equation in Eq. (\ref{eqsup:continuity}) straightforwardly translates into $\partial_\mu\mc J^\mu = 0$, since 
\begin{align}
    \begin{split}
        \partial_t\rho_v + d\star j_v &= \partial_t(d\xi^1d\xi^2 \mc J^0) + d\xi^1d\xi^2 \partial_i \mc J^i \\
        &= d\xi^1d\xi^2(\partial_t \mc J^0 + \partial_i \mc J^i)\\
        &= d\xi^1d\xi^2(\partial_\mu \mc J^\mu) = 0.
    \end{split}
\end{align}
We note that the topological continuity equation is simply a reformulation of the generalized Stokes' theorem. Importantly, in deriving this continuity equation we have not specified any Lagrangian. Therefore, this continuity equation and conservation of the vorticity density is robust and will not be spoiled by any disorder or structural symmetries associated with the Lagrangian.


We can directly check that the continuity $\partial_\mu \mc J^\mu = 0$ is satisfied for $\mc J^\mu$ in Eq. (\ref{eqsup:vorticity_current}). Let us now allow the membrane to be dynamic and demonstrate that the continuity equation is still satisfied even when the membrane can fluctuate. Doing so, the ``electric" component of the field strength tensor is now nonzero, and $\*e_1$, $\*e_2$, and $\*n$ may be time dependent. $\*m$ may also fluctuate in both orientation as well as magnitude.  We have that 
\begin{equation}
    \begin{aligned}
        \partial_\mu \mc J^\mu &= \partial_\mu \left\{\frac{\epsilon^{\mu\nu\rho}}{2\pi}\left[\*n\cdot(\nabla_\nu\*m\times\nabla_\rho\*m) - \frac{1}{2}\mc F_{\nu\rho}\*m_\|^2\right]\right\}
    \end{aligned}
\end{equation}
Let us simplify the first and second terms in the brackets separately by expanding the covariant derivative and expressing the terms with the regular derivative $\partial_\mu$ and the gauge potential $\mc A_\mu$. For the first term, we have
\begin{equation}
    \begin{aligned}
        \frac{\epsilon^{\mu\nu\rho}}{2\pi}\left[\*n\cdot(\nabla_\nu\*m\times\nabla_\rho\*m)\right]= \frac{\epsilon^{\mu\nu\rho}}{2\pi} \{(\partial_\nu m^a)(\partial_\rho m^b)\*n\cdot(\*e_a\times\*e_b) &- \mc A_\rho(\partial_\nu m^a)m^b\*n\cdot[\*e_a\times(\*n\times\*e_b)]\\
        &-\mc A_\nu (\partial_\rho m^b)m^a\*n\cdot[(\*n\times\*e_a)\times\*e_b]\}\\
        = \frac{\epsilon^{\mu\nu\rho}}{2\pi} \{(\partial_\nu m^a)(\partial_\rho m^b)\*n\cdot(\*e_a\times\*e_b) &- 2\mc A_\rho(\partial_\nu m^a)m^b\*n\cdot[\*e_a\times(\*n\times\*e_b)]\\
        = \frac{\epsilon^{\mu\nu\rho}}{2\pi} \{(\partial_\nu m^a)(\partial_\rho m^b)\*n\cdot(\*e_a\times\*e_b) &- \mc A_\rho\partial_\nu (\*m_\|^2)\}
    \end{aligned}
\end{equation}
Taking the divergence of the simplified first term,  we see that
\begin{equation}
    \frac{\epsilon^{\mu\nu\rho}}{2\pi}\partial_\mu\left[\*n\cdot(\nabla_\nu\*m\times\nabla_\rho\*m)\right] = -\frac{\epsilon^{\mu\nu\rho}}{2\pi}(\partial_\mu \mc A_\rho)\partial_\nu(\*m_\|^2) = \frac{\epsilon^{\mu\nu\rho}}{2\pi} \frac{\mc F_{\nu\rho}}{2}\partial_\mu(\*m_\|^2).
\end{equation}
Here, we have used the fact that the contraction of the antisymmetric Levi-Civita symbol $\epsilon^{\mu\nu\rho}$ with the symmetric terms $\partial_\mu\partial_\nu m^a$ and $\partial_\mu\partial_\rho m^b$ is zero, and that $\*n\cdot(\*e_a\times\*e_b)$ is a constant. Next, we take the divergence of the second term to get 
\begin{equation}
    \frac{\epsilon^{\mu\nu\rho}}{2\pi}\partial_\mu\left(-\frac{1}{2}\mc F_{\nu\rho}\*m_\|^2\right) = -\frac{\epsilon^{\mu\nu\rho}}{2\pi}\frac{\mc F_{\nu\rho}}{2}\partial_\mu(\*m_\|^2),
\end{equation}
using the identity that $\epsilon^{\mu\nu\rho}\partial_\mu\mc F_{\nu\rho}=0$. Since the two terms sum to zero, it is indeed true that 
\begin{equation}
    \partial_\mu \mc J^\mu = 0.
\end{equation}

It is interesting to note that this formalism can be extended to higher dimensions. Let us consider an $n$-dimensional orientable manifold $\mc M$ embedded in an $(n+1)$-dimensional space. On the manifold, there is an $(n+1)$-dimensional order parameter texture. At every point on $\mc M$, suppose there exists a local hard axis normal to $\mc M$, which follows the curvature of $\mc M$ and makes it preferential for the order parameter to lie in $\mc M$. 

Let us define, as a function of the order parameter field, an $(n-1)$-form $\Tilde{\rho}_w$. Suppose that in the strong ``easy space" limit in which the order parameter lies fully in $\mc M$, we can construct $\Tilde{\rho}_w$ such that integrating it over the $(n-1)$-dimensional boundary yields a topological charge. In such a case, we can derive a topological continuity equation in curved space. The density $n$-form $\Tilde{\rho}_v$ and the flux 1-form $\Tilde{j}_v$ can be derived from $\Tilde{\rho}_w$ by invoking the generalized Stokes' theorem. We define $\Tilde{\rho}_v$ as the exterior derivative of $\Tilde{\rho}_w$, yielding $\Tilde{\rho}_v = d\Tilde{\rho}_w$. Applying the Hodge star map \cite{stone09,nakahara18} and taking the time derivative of both sides, $\Tilde{\rho}_v = d\Tilde{\rho}_w$ is recast as the continuity equation $\star \partial_t\Tilde{\rho_v} + \star d \star \Tilde{j_v}=0$, where $\Tilde{j}_v =(-1)^n \star \partial_t\Tilde{\rho}_w$.

The known topological continuity equation for hedgehogs in three-dimensional magnets \cite{zouPRL20} can be understood as a limiting case in which the 3-dimensional space is flat, and the order parameter texture lies entirely within the ``easy space."

\subsection{(iii) Torsion of a curve/helix}

Torsion is a geometric property of a curve embedded in $\mb R^3$ which can be used to reduce symmetries and enable vortex dynamics. In this section, we will calculate the torsion of a uniform helix, which is used in the setup discussed in the main text. A uniform helix can generally be described by 
\( \*r(\ell)=\left(r\cos\frac{\omega\ell}{c}, r \sin \frac{\omega\ell}{c}, \frac{b}{c}\ell   \right).   \)
Let us assume $r,\omega, c>0$, and then the sign of $b$ determines the chirality of the helix. $\text{sgn}(b)=\pm1$ corresponds to positive or negative chirality. The velocity is given by: 
\( \*{v}=\dv{\*r}{\ell}=\left( -\frac{r\omega}{c}\sin \frac{\omega\ell}{c}, \frac{r\omega}{c} \cos\frac{\omega\ell}{c}, \frac{b}{c}  \right),   \)
which is not unity. Imposing the constraint that $|\* v|=1$, we obtain $c^2=r^2\omega^2+b^2$. Therefore, we should take $c=\sqrt{r^2\omega^2+b^2}$ when using the arc length to parameterize the helix. The acceleration is given by
\( \dv{\*v}{\ell}=\left( -\frac{r\omega^2}{c^2}\cos\frac{\omega\ell}{c}, - \frac{r\omega^2}{c^2}\sin\frac{\omega\ell}{c} , 0  \right).  \)
The curvature is its magnitude, 
\( \kappa(\ell)= \frac{r\omega^2}{c^2},  \)
which is independent of $\ell$. The normal vector is then 
\( \*a(\ell)=\frac{d\*v/d\ell}{\kappa}=\left( -\cos\frac{\omega\ell}{c}, -\sin\frac{\omega\ell}{c}, 0   \right). \)
We can then determine the bi-normal vector 
\( \*b(\ell)=\*v\times \*a=\left( \frac{b}{c}\sin\frac{\omega\ell}{c}, - \frac{b}{c}\cos\frac{\omega\ell}{c}, \frac{r\omega}{c}  \right) . \)
To determine the torsion, we evaluate the changing rate of this vector, which yields
\( \dv{\*b}{\ell} =\left( \frac{b\omega}{c^2}\cos\frac{\omega\ell}{c}, \frac{b\omega}{c^2}\sin\frac{\omega\ell}{c}, 0  \right).  \)
We recall the definition of the torsion, $\frac{d\*b}{d\ell}=-\mathfrak{T}\*a$, from which we obtain the torsion 
\( \mathfrak{T}(\ell)=\frac{b\omega}{c^2}= \frac{b\omega}{r^2\omega^2 +b^2}.   \)
We remark that both curvature and torsion of a helix are independent of the point $\ell$, as one may expect. It is important to notice that the torsion changes the sign when we switch the chirality of the helix. Finally, the torsion can be written in terms of the helix angle $\theta$. To see this, we first note that 
\begin{equation}
    \sin\theta = \frac{b}{\sqrt{r^2\omega^2+b^2}},~~~\cos\theta = \frac{r\omega}{\sqrt{r^2\omega^2+b^2}}.
\end{equation}
The torsion can thus be written as
\begin{equation}
    \mf T = \frac{\sin(2\theta)}{2r}.\label{eq:torsion}
\end{equation}

\subsection{(iv) Construction of phenomenological torque}

In this section, we phenomenologically construct a torque per unit length that acts on the magnetic film by pumping vorticity transverse to the electric wire into the bulk of the magnet. A possible torque that achieves this would be a torque that performs work on the magnet proportional to 
\begin{align}
    \delta \mc Q&=\frac{1}{2\pi} \int\limits_{\text{wire}}d\ell \,\*n\cdot\left[(\delta\*m\times\nabla_\ell\*m)+(\*m\times\nabla_\ell \delta \*m)\right],\\
    &=\frac{1}{2\pi} \int\limits_{\text{wire}}d\ell \left\{2(\nabla_\ell\*m\times \*n)\cdot\delta\*m+\partial_\ell\left[\*n\cdot(\*m\times \delta \*m)\right]\right\} \label{eq:bounadrywinding},
\end{align}
where we varied Eq.~(\ref{eq:tangent_winding}) with respect to $\vb{m}$. In the second line, we used the identity from Eq~\eqref{eqsup:identity}. In the following, we omit the second term in Eq.~\eqref{eq:bounadrywinding}, which injects winding from the boundary. 
Using this, we can reverse-engineer the torque $\bs{\mc \tau}$  to be
\begin{equation}
    \begin{aligned}
    \bs{\tau} &=\zeta j\mf T \,\*m \times \frac{\delta \mc Q}{\delta \*m}\\
   &=\frac{\zeta j\mf T}{\pi} \*m\times(\nabla_\ell\*m \times \*n)\\
    &= \frac{\zeta j}{\pi}(\mf T\*n\cdot\*m)\left[\nabla_\ell\*m + \*n\partial_\ell(\*n\cdot\*m)\right].\label{eq:torqueconst}
    \end{aligned}
\end{equation}
The coefficient of proportionality is determined by requiring that an electrical current $j$ is used to bias the vortex injection, the work done on the magnetic texture is proportional to the torsion of the wire $\mf T$ (to satisfy symmetry requirements), and $\zeta$ is a phenomenological coefficient that may depend on the strength of the spin-orbit coupling. 
In the last line, we rewrite the torque into a form similar to the torque in flat geometries~\cite{zouPRB19,jonesPRB20}. 

We can verify that the work done on the magnetic texture by the torque is
\begin{equation}
    \begin{aligned}
        \delta W &= \int d\ell\,dt\,\bs \tau \cdot (\*m\times\partial_t\*m)\\
        &= \frac{\zeta \mf Tj}{\pi}\int d\ell\,dt\,\*n\cdot (\partial_t\*m\times \nabla_\ell\*m)\\
        &= \zeta\mf T j\,\delta\mc Q,
    \end{aligned}
\end{equation}
where in the second step, we inserted the second line of Eq.~\eqref{eq:torqueconst} and used that $\abs{\vb{m}}=1$.
In the last step, we identify the change in vortex charge $\delta \mc Q$ as the integrated vorticity current across the wire, see Eq.~\eqref{eqsup:vorticity_current}. 

Another approach to construct this torque is to first consider the simplest torque (i.e. to leading order in spatial gradients) that pumps vorticity in flat geometries,
\begin{equation}
    \bs \tau_{\text{flat}} = \eta (\*M\cdot\*m)(j\partial_\ell) \*m.
\end{equation}
This torque was used in Refs. \cite{zouPRB19} and \cite{jonesPRB20}. Using $\bs \tau_{\text{flat}}$ as a blueprint, we substitute $\*M$ with $\mf T\*n$, promote the normal derivative of $\*m$  to the covariant derivative of $\*m$, and project out components of the torque parallel to $\*m$. By doing so, we arrive at the torque proposed in our manuscript.  While the spatial properties of $\*M$ and $\mf T\*n$ are equal, they differ under time reversal, which makes our torque dissipative. 

\subsection{(v) Estimation of effective vortex circuit elements}

The section ``Vortex circuit elements" in the main text introduces the effective vortex resistance $R_v$, the effective winding capacitance $C_v$, and the effective drag $\mf R$ for the setup of the cylindrical magnetic membrane wrapped with a wire. These circuit elements are then used to construct effective topological circuits. In this section, we elucidate how expressions for $\mf R$, $R_v$ and $C_v$ may be estimated. First, let us describe the setup. A magnetic insulating membrane of thickness $h$ and length $l_m$ wraps around a cylindrical insulating core. A metal wire of width $w$ and thickness $\delta$ is wrapped around the cylindrical magnetic membrane of radius $r$ as a uniform helix with helix angle $\theta$. 

\begin{figure}[b]
    \centering
    \includegraphics[width = 0.6\linewidth]{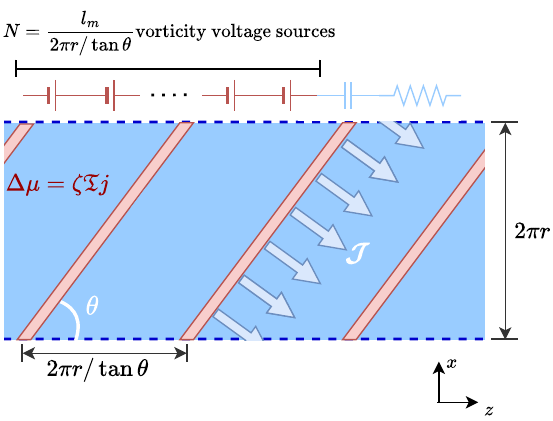}
    \caption{An ``unrolled" depiction of a cylindrical magnetic membrane. A metal wire is wrapped as a uniform helix around the magnetic cylinder. The wire translates into evenly spaced metal stripes in the flat depiction. The setup can be understood as a series $R_vC_v$ circuit of $N$ voltage sources connected in series with the vortex resistance $R_v$ and the winding capacitance $C_v$. Each voltage source supplies a motive force $\Delta \mu$ to the effective vortex circuit when an electric current flow is induced in the wire.}
    \label{fig:suppfig2}
\end{figure}

Although the magnetic membrane is curved, to estimate the vortex circuit elements, it is useful to ``unroll" the cylinder and consider vortex dynamics on a flat magnetic membrane. This is depicted in Fig.~\ref{fig:suppfig2}. 
Here, we see that the magnetic membrane translates to a rectangle of width $2\pi r$ and length $l_m$. The helically wrapped wire translates into metallic stripes across the magnetic rectangle at angle $\theta$ and spaced out at distances of $2\pi r/\tan\theta$, which is the pitch of the helix. When an electric current flow is induced in the metallic wire, the electric current exerts a torque on the magnetic texture which acts to bias vorticity injection transverse to the wire. The magnetic system responds by exhibiting a vortex resistance, which can stem from Gilbert damping, vortex-antivortex collisions, and defects, and a winding capacitance, which arises due to the magnetic stiffness. The system depicted in Fig.~(\ref{fig:suppfig2}) shows that the setup of the magnetic cylinder wrapped with a wire is effectively a series $R_vC_v$ circuit of $N$ vorticity voltage sources, a vortex resistor, and a winding capacitor, where 
\begin{equation}
    N = \frac{l_m\tan\theta}{2\pi r}
\end{equation}
is the number of metallic stripes. Let us first derive the effective drag coefficient $\mf R$.  When electric current flow is induced in the metal wire, each metal stripe supplies a vortex motive force
\begin{equation}
    \Delta \mu = \zeta \mf T j 
\end{equation}
to the effective vortex circuit. Here, $\zeta$ is the phenomenological parameter characterizing the charge-vortex coupling, $\mf T$ is the torsion, and $j$ is the electric current density. The vortex chemical potential is taken from Eq.~(7) in the main text. The total supplied motive force is thus
\begin{equation}
    \mc V = N\Delta\mu = \frac{\zeta \mf Tl_m\tan\theta}{2\pi r}j = \frac{\zeta \mf T}{w\delta}\frac{l_m}{2\pi r}I\tan\theta, 
\end{equation}
where $I$ is the electric current, $w$ is the width of the wire, and $\delta$ is the thickness of the wire. The vortex motive force can be thought of as the drag force exerted by the electric current on vortices, where the drag coefficient is
\begin{equation}
    \mf R \equiv \frac{\mc V}{I}= \frac{\zeta \mf T}{w\delta}\frac{l_m}{2\pi r}\tan\theta. \label{eq:seebeck}
\end{equation}
This is the same as Eq.~(8) in the main text. Next, the vortex resistance can be estimated using Ohm's law. The component of the vortex flux $\mc J$ which flows along the $z$ direction is related to the supplied motive force by
\begin{equation}
    \mc J \sin\theta= \sigma_v\frac{\mc V}{l_m},
\end{equation}
where $\sigma_v$ is the vortex conductivity. The vortex current $I_v$ which flows along $z$ and builds up winding azimuthally is
\begin{equation}
    I_v =  2\pi r \mc J\sin\theta.
\end{equation}
Thus, we define the vortex resistance as
\begin{equation}
    R_v \equiv \frac{\mc V}{I_v} = \frac{1}{\sigma_v}\frac{l_m}{2\pi r}.
\end{equation}
Finally, we estimate the winding capacitance $C_v$. We assume that in the steady state, the magnetic texture is loaded with vortex charge $\mc Q_s$, and the azimuthal winding is homogenous, such that the winding density along a loop around the cylinder at fixed $z$ is constant and may be given by
\begin{equation}
    \frac{1}{2\pi}\*m_\|^2 D_\ell \varphi = \frac{\mc Q_s}{2\pi r},
\end{equation}
where $\ell$ is the arclength of the loop. Anticipating that the energy stored on the magnetic texture will be related to $C_v$ by $\mc E = \mc Q_s^2/2C_v$, we can roughly estimate the steady-state energy that arises due to the magnetic winding by
\begin{equation}
    \begin{aligned}
        \mc E &\sim \frac{hA}{2}\int dS (\*m_\|^2D_\ell\varphi)^2\\
        &= \frac{hA}{2}\int dS  \left(\frac{\mc Q_s}{r}\right)^2\\
        &= \pi hA\mc Q_s^2l_m/r.
    \end{aligned}
\end{equation}
Equating this with $\mc Q_s^2/2C_v$, we find that the winding capacitance may be estimated as
\begin{equation}
    C_v = \frac{1}{A}\frac{r}{2\pi h l_m}.
\end{equation}
The charging efficiency can be given by
\begin{align}  \eta_c=\frac{{C_v}V_v(t)^2/2}{\int_0^t\mathrm{d} t^\prime  V(t^\prime)I_0}&=\frac{\xi}{2}\frac{(1-e^{-t/\tau})^2}{t/\tau-\xi (1-e^{-t/\tau})},
\end{align}
where we solved for $V(t)=R I_0 \left(1-\xi e^{-t/\tau}\right)$ and $V_v(t)=-\mf{R} I_0 \left(1-e^{-t/\tau}\right)$ using a constant current $I_0$ and $V_v(0)=0$.
It fulfills the limit $\lim_{t\rightarrow 0}\lim_{\xi\rightarrow 1}\eta_c=1$, where the order of limits is important. Using as a charging time $t=\tau$, we obtain the result of the main paper. The discharging efficiency is given by
\begin{align}
    \eta_d=\frac{\int_0^t\mathrm{d} t^\prime R_{L} I^2(t^\prime)}{\int_0^t\mathrm{d} t^\prime  V_v(t^\prime)I_v(t^\prime)}=\frac{1-\gamma}{1+(\mathcal Z_v\gamma)^{-1}},
\end{align}
where 
$\gamma=R_L/(R+R_L)$ and we used that $V(t)=0$ and $V_v(0)>0$.

To optimize the efficiency of the battery, we have to maximize the vortexoelectric figure of merit ${\mathcal Z}_v$ which is a monotonic function of $\xi={\mf{R}}^2/RR_v$. The electrical resistance $R$ is estimated in accordance with Drude model by taking $R = \rho l_m/w\delta\cos\theta$. Here, $l_m/\cos\theta$ is the length of the metal wire and $w\delta$ is the cross-sectional area. $\rho$ is the electrical resistivity, which is estimated to be $\rho\sim \hbar\lambda_F^2/e^2d$, with $d$ the mean free path, $\lambda_F$ the Fermi wavelength and $e$ the positive electric charge. The vortex resistance $R_v$ can be estimated near the Curie temperature, in which the vortex conductivity is given by $\sigma_v = \rho_v D/k_BT_c$ via the Einstein relation. Close to the transition temperature $T_c\sim J/k_B$, the vortex density $\rho_v\sim 1/a^2$ varies on the order of the lattice spacing, and the diffusion coefficient can be estimated by $D\sim Ja^3/\hbar h$. The estimate for $\mf R$ is obtained by using Eq.~\eqref{eq:seebeck} and Eq.~\eqref{eq:torsion}, in conjunction with the estimate for the coupling coefficient $\zeta \sim \hbar w\lambda_F^2/e$. Altogether, we find that 
\begin{subequations}
    \begin{align}
        R \,\,\,&\sim \frac{l_m}{w\delta \cos\theta}\frac{\hbar \lambda_F^2}{e^2d}\\
        R_v &\sim \frac{hl_m}{r}\frac{\hbar}{a}\\
        \mf R \,\,&\sim \frac{l_m \sin^2\theta}{r^2\delta}\frac{\hbar \lambda_F^2}{e}
    \end{align}
\end{subequations}
From these estimates, we find that the dependence of $\xi$ on the helix angle $\theta$ is given by $\xi\propto \sin^4\theta\cos\theta$, thus being maximized by $\theta \approx 63^\circ$.

\bibliography{ms.bib}